\documentclass[pbalance=true, screen, nonacm, sigconf]{acmart}

\usepackage[defaultlines=3,all]{nowidow}
\usepackage{siunitx}
\usepackage[dvipsnames]{xcolor}
\usepackage{csquotes}

\usepackage{minted}

\usepackage[capitalise,noabbrev]{cleveref}
\crefformat{section}{\S#2#1#3} 
\crefformat{subsection}{\S#2#1#3}
\crefformat{subsubsection}{\S#2#1#3}
\crefrangeformat{section}{\S#3#1#4 to~\S#5#2#6}
\crefrangeformat{subsection}{\S#3#1#4 to~\S#5#2#6}
\crefrangeformat{subsubsection}{\S#3#1#4 to~\S#5#2#6}

\newcommand{\name}{\textsc{Beaver}}
\newcommand{\faas}{Function-as-a-Serivce}

\title[Serverless Abstractions for Short-Running, Lightweight Streams]{Serverless Abstractions for Short-Running, Lightweight Streams}

\author{Natalie Carl}
\affiliation{
    \institution{Technische Universität Berlin}
    \city{Berlin}
    \country{Germany}
}
\email{nc@3s.tu-berlin.de}
\orcid{0009-0000-5991-9255}

\author{Niklas Kowallik}
\affiliation{
    \institution{Technische Universität Berlin}
    \city{Berlin}
    \country{Germany}}
\email{nk@3s.tu-berlin.de}
\orcid{0009-0006-9839-1864}

\author{Constantin Stahl}
\affiliation{
    \institution{Technische Universität Berlin}
    \city{Berlin}
    \country{Germany}
}
\email{cs@3s.tu-berlin.de}
\orcid{}

\author{Trever Schirmer}
\affiliation{
    \institution{Technische Universität Berlin}
    \city{Berlin}
    \country{Germany}}
\email{ts@3s.tu-berlin.de}
\orcid{0000-0001-9277-3032}

\author{Tobias Pfandzelter}
\affiliation{
    \institution{Technische Universität Berlin}
    \city{Berlin}
    \country{Germany}}
\email{tp@3s.tu-berlin.de}
\orcid{0000-0002-7868-8613}

\author{David Bermbach}
\affiliation{
    \institution{Technische Universität Berlin}
    \city{Berlin}
    \country{Germany}
}
\email{db@3s.tu-berlin.de}
\orcid{0000-0002-7524-3256}

\begin{abstract}
    Serverless computing and stream processing represent two dominant paradigms for event-driven data processing, yet both make assumptions that render them inefficient for short-running, lightweight, and unpredictable streams that require stateful processing.
    We propose \emph{stream functions} as a novel extension of the \faas{} model that treat short streams as the unit of execution, state, and scaling.
    Stream functions process streams via an iterator-based interface, enabling seamless inter-event logic while retaining the elasticity and scale-to-zero capabilities offered by serverless platforms.
    Our evaluation shows that stream functions reduce the processing overhead by \textasciitilde{}\SI{99}{\percent} compared to a mature stream processing engine in a video-processing use case. 
    By providing comparable performance to serverless functions with stream semantics, stream functions provide an effective and efficient abstractions for a class of workloads underserved by existing models.
\end{abstract}

\begin{document}

    \maketitle

    \section{Introduction}\label{sec:intro}

\faas{} (FaaS) is an event-based cloud computing model in which developers write small, stateless functions to process individual, discrete events.
A serverless computing platform handles all aspects of the function lifecycle: from low-level resource management all the way up to the function execution environment, function invocation, and function instance scheduling~\cite{jonas2019berkeley}.
In turn, developers can focus primarily on their application logic and large, constant infrastructure expenses are omitted through pay-by-use billing.
This paradigm has been applied to building REST APIs, processing data from IoT devices, responding to database updates, or running regularly scheduled tasks and cron jobs, among others~\cite{eismann2022stateofserverless}. 

Arguably, the ability to effortlessly scale is FaaS' primary selling point.
On the one hand, statelessness gives rise to intrinsic elasticity: Serverless platforms maintain dynamic pools of function instances that are used interchangeably for processing incoming requests.
Function instances are spawned or removed according to load and configuration --- in turn reducing idle resource use and cost by scaling to zero.
The resulting ability to scale elastically allows FaaS applications to process highly bursty workloads in near real-time~\cite{jonas2017occupy}.

This, however, does not come for free.
There are no guarantees for two requests to be processed by the same instance because serverless platforms keep a tight grip on scheduling.
Consequently, keeping state requires either external storage or direct communication between function instances.
The former is slow and costly, and public serverless platforms lack first-class mechanisms for the latter~\cite{copik2023fmi}.
This limits FaaS applications to use cases where computation across events is infrequent or extraneous altogether.
More specifically, serverless functions are unfit for handling continuous streams with inter-event processing logic.

Although different approaches for stateful serverless computing have been proposed, these are mostly unable or not intended to compete with full-blown stream processing engines designed for that very purpose~\cite{pfandzelter2022streamingcost}.
Stateful serverless computing platforms equip functions with key--value stores to keep state across function invocations~\cite{sreekanti2020cloudburst}.
This, however, still requires a serverless platform to schedule, invoke, and scale function instances in response to small parts of the stream instead of keeping a function alive while a network connection is open, through which data can be sent continuously for the duration of the stream.
In contrast, stream processing engines are designed for processing long-running, continuous streams with high, albeit relatively steady throughput and low latency. 
Developers express computation as graphs of operators connected by data streams, which are deployed ahead of time and remain active regardless of data arrival~\cite{google2015dataflow}. 
This model enables efficient inter-event processing, ordering guarantees, and stateful computation by amortizing initialization and resource allocation costs over the lifetime of a continuously running pipeline. 
However, as we will demonstrate, these assumptions limit the flexibility and elasticity that serverless applications provide.
In particular, stream processing engines are inefficient for short-lived or sporadic streams, as startup overhead and always-on resource allocation dominate execution time when stream arrival is unpredictable or infrequent.

In this paper, we argue that a class of streaming applications exists for which neither current \faas{} nor stream processing approaches are well suited.
In particular, this class is defined by four characteristics.
(1) Streams are \textbf{short-running}, in the order of seconds or minutes.
(2) They are \textbf{lightweight}, with throughput in the megabytes per second.
(3) Their time of arrival is \textbf{unpredictable}.
(4) The application must have the ability to process streams in a \textbf{stateful} manner.
Together, these characteristics demonstrate where current FaaS offerings and stream processing engines fall short.
While serverless functions handle short-running, lightweight, unpredictable requests with ease, they struggle when inputs are continuous and require stateful processing~\cite{pfandzelter2022streamingcost}.
In contrast, stream processing engines excel at continuous tasks but struggle with quickly scaling to zero and using resources only when necessary~\cite{song2023sponge}.
This leaves a gap between \faas{} and stream processing for which no appropriate programming abstractions exist.

To address this, we outline different examples for such applications (\cref{sec:examples}), 
compare serverless and stream processing abstractions (\cref{sec:state}),
propose stream functions as an alternative by extending the \faas{} model (\cref{sec:approach}), 
and compare our proof-of-concept implementation against different stream processing and serverless alternatives (\cref{sec:eval}).

    \section{Motivating Examples}\label{sec:examples}

The purpose of the following examples is to illustrate the need for a suitable programming abstraction for efficiently processing short-running, ephemeral streams.
While each example \emph{could} be addressed differently using existing technology, we point out different shortcomings to motivate the need for novel abstractions.

\paragraph{Example 1: On-demand video processing}
Consider on-demand video pre-processing pipelines that are triggered by user interaction on social media platforms.
In contrast to traditional media pipelines, which are designed to process large volumes of continuously arriving data, these workloads operate on short video streams with durations of only a few seconds and modest throughput.
The arrival of such streams is inherently unpredictable and depends on user behavior or external triggers.
Provisioning continuously running video processing pipelines to handle these sporadic workloads leads to low utilization and unnecessary cost given likely daily or weekly usage patterns.
At the same time, processing individual frames using stateless serverless functions complicates inter-frame operations, as these require state to be preserved across frames.
Latency is often a first-class concern: Pre-processing is frequently on the critical path for downstream tasks such as content delivery, analytics, or further processing stages.
The system must therefore rapidly instantiate processing logic, maintain per-stream state for the duration of the video segment, and release resources immediately after completion.
Existing stream processing engines assume long-lived pipelines and amortize startup costs over time, while serverless platforms lack abstractions for binding execution and state to the lifetime of a short video stream.
This makes on-demand video pre-processing pipelines a representative example of workloads that fall between traditional stream processing and serverless computing.
In \cref{sec:eval}, we will use a simplified version of this use case for our evaluation.

\paragraph{Example 2: Natural disaster monitoring}
Scientific monitoring systems observe natural hazards such as earthquakes, floods, or volcanic activity using distributed sensor networks, including seismometers, river gauges, and weather stations.
While data is collected continuously, detailed analysis is typically triggered only when specific conditions are met, such as threshold exceedances or anomaly detection.
When triggered, a short burst of sensor data --- spanning seconds to minutes --- is streamed for immediate processing.
This processing requires maintaining state over the duration of the stream, for example to compute rolling statistics, track recent baselines, or correlate measurements across sensors.
The arrival time of such streams is inherently unpredictable, given the rarity of such physical events.
Continuously running analytics pipelines for all sensors is inefficient, while processing individual measurements with stateless serverless functions is infeasible. 
A suitable system must therefore rapidly scale up to process bounded streams with per-stream state and scale down immediately once processing completes.
Existing stream processing engines and serverless platforms each fall short of these requirements, motivating the need for alternative programming abstractions.

\paragraph{Example 3: Interactive data analysis}
In interactive analytics systems, users frequently issue queries that trigger short bursts of data to be streamed for immediate processing, such as exploratory aggregations or transformations.
These streams are short-lived, their time of arrival is unpredictable, they often require maintaining intermediate state across records, and they need to be processed as quickly as possible.
Provisioning long-running analytics pipelines for such sporadic interactions is inefficient, while event-at-a-time serverless execution complicates stateful processing. 
Supporting interactive analytics therefore requires processing short, stateful streams with rapid startup and immediate scale-down.

    \section{Serverless and Stream Processing Models}\label{sec:state}

Modern cloud platforms offer FaaS and stream processing as the two dominant programming models for processing event-driven data.
While both are widely used in practice, they make fundamentally different assumptions about execution, state, and scaling.

\paragraph{Serverless}
FaaS platforms expose an event-driven programming model in which functions are invoked independently for each incoming event.
The platform manages scheduling, scaling, and isolation, allowing applications to elastically scale based on demand.
Function instances are stateless by default, and there is no guarantee that related events are processed by the same instance.
As a result, serverless platforms scale events, not streams.
Ordering between events is not preserved automatically, and any inter-event logic requires external state or explicit communication between function instances.
While this design enables fine-grained elasticity and efficient resource usage for independent requests, it complicates applications whose computation naturally spans multiple related events, such as continuous or session-based data streams.

\paragraph{Stream processing}
Stream processing engines follow a pipeline-oriented programming model in which developers define long-running graphs of operators connected by data streams.
Inter-event logic is expressed using operators and windowing constructs, which group events based on time or other criteria.
These systems scale by parallelizing operators and provide strong processing guarantees through checkpointing and replay.
Depending on the platforms, operators are either specified through declarative, SQL-like languages or by implementing custom processing logic in a high-level programming language.
This model assumes that pipelines are deployed ahead of time and remain active regardless of data arrival.
Startup costs and resource usage are amortized over long-running workloads, making stream processing engines well suited for sustained, high-throughput streams.
However, as we will argue, these assumptions make them inefficient for short-lived or sporadic streams, where initialization overhead dominates execution time.

\paragraph{Bridging attempts}
Several systems attempt to combine serverless elasticity with streaming semantics, for example by adding stateful storage to serverless functions or by deploying streaming frameworks in an elastic, on-demand manner.
While these approaches reduce some overheads, they typically retain the core abstractions of either event-at-a-time serverless execution or pipeline-based stream processing.
Consequently, existing models lack first-class support for treating a short-lived stream as the unit of execution, scaling, and state lifetime.
This motivates the need for a programming abstraction that combines streaming semantics with the elastic, pay-per-use characteristics of serverless platforms.

    \section{Stream Functions}\label{sec:approach}

We propose stream functions as a novel extension of the \faas{} model.
We believe they offer the most natural programming abstraction for processing highly fluctuating amounts of short, lightweight streams in a resource-efficient manner.

\subsection{Programming model: Iterator in, iterator out}

In serverless computing, the function signature serves as the interface between platform and user code.
This dictates the programming abstraction.
Each function receives an individual event as input, such as an HTTP request or cloud event, and optionally returns a response.
In contrast, a stream function receives an iterator as input and optionally returns an iterator.
The function code then determines how the entire stream in processed, compared to specifying logic for individual events only.
\cref{fig:streamfn} illustrates how stream functions look in the Go programming language.
A developer supplies the function Fn, which is deployed to the serverless platform.
The stream consists of events that comprise a payload of bytes and metadata, such as a timestamp and other metadata.
By iterating over the input stream, events are processed, and inter-event logic can be expressed as desired.
In this example, developers call the yield function in order to output new events to other processing steps.

\begin{listing}[h]
    \begin{verbatim}
type Event struct {
    Headers map[string]any
    Payload []byte
}

func Fn(stream iter.Seq[Event]) iter.Seq[Event] {
    return func(yield func(Event) bool) {
        for e := range stream {
            // ...
        }
    }
}
    \end{verbatim}
    \caption{
        Go skeleton of a stream function and type definition.
        The user-supplied function Fn iterates over all elements of the streams, returning outputs by calling the yield function on demand.
    }
    \label{fig:streamfn}
\end{listing}

Similarly to traditional FaaS functions, the processing logic is straightforward to implement and functions are portable.
During deployment, users specify an output for processed events.
Chaining multiple stream functions into a workflow enables building arbitrarily complex processing pipelines.
Each stream function corresponds to an individual processing step in a workflow.
Consequently, no optimization of an application graph is performed --- in turn, new stream function instances can be spun up or destroyed quickly, independently of other processing steps.
Parallel processing of multiple streams is analogous to processing multiple request with multiple function instances.
Akin to traditional FaaS functions, instances can be isolated through processes, containers, or µVMs, among others~\cite{agache2020firecracker,young2019gvisor}.
Moreover, iterators are first-class components of most high-level programming languages.
While we use Go for illustration, this concept maps cleanly onto other languages popular for serverless computing, enabling seamless adoption.

\subsection{Function instances in detail}

\begin{figure}[h]
    \includegraphics{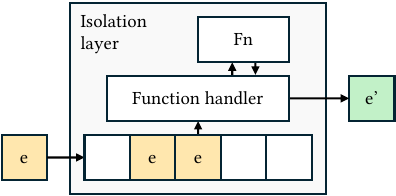}
    \caption{
        Components of a stream function instance.
        Events are buffered before being passed to the function by the function handler.
        Transformed events are emitted according to configuration.
    }
    \label{fig:function}
\end{figure}

As shown in \cref{fig:function}, stream function instances consist of three components: (1) the user-supplied function, (2) a FIFO message buffer, and (3) a function handler transferring inputs from the message buffer to the user function and handling optional outputs.
The function handler and user-supplied function exist in the same process to minimize the overhead of passing messages between them.
Its purpose is to provide an abstraction over the lower-level handling of the stream itself and limit user access to the buffer.
The queue serves as a small buffer and is intended only for additional storage during initialization and to provide leeway when processing events temporarily exceeds the mean time between arrivals.

All components are wrapped into a container for isolation between function instances.
Note that we intentionally omit details on serverless platform design and are only concerned with the programming model and structure of the function instances.
Hence, depending on the implementation, different messaging and isolation mechanism can be used interchangeably.

\subsection{The function lifecycle}\label{subsec:lifecycle}

The lifecycle of stream function instances comprises three phases.
(1)~As soon as a stream producer opens a connection to the platform, the control plane creates a new instance, the message buffer is initialized, and the function handler starts equipped with the user function.
(2)~For the duration of the stream, the function instance continues processing and (optionally) emits processed events.
Importantly, the user function only ever receives a single function call during initialization.
Instead, the function handler supplies the user function with new events through the iterator it is called with.
(3)~As soon as the network connection between the stream producer and the platform is closed, the instance is shut down and memory allocated to the buffer is freed.
Note that for improved cold start latency, platform implementations for stream functions may keep instances warm for later reuse.
By default, we immediately scale to zero after a stream has ended.

    \section{Evaluation}\label{sec:eval}

The goal of stream functions is to drastically reduce the overhead of system initialization for processing short, lightweight streams.
To this end, we compare our proof-of-concept implementation against three alternatives as a baseline.

\subsection{A video processing use case}

We implement a load generator that generates a stream of images at 10 frames per second with a 160 $\times$ 120 resolution.
The stream consists of a repeating series of 20 randomly generated and distinct images that are not cached by the processing step.
For each system, we provide an adapter that transfers the stream using the preferred protocol.
A single processing step converts each frame to grayscale by setting the RGB values of each pixel to their average.

\paragraph{Prototype and baselines}
We implement our proof-of-concept prototype \name{}, which we publish as open source.
It consists of a control plane, which accepts incoming network connections and handles the function lifecycle (\cref{subsec:lifecycle}), and a function instance for video processing as outlined above.
Together with the message buffer, for which we use \href{https://nats.io/}{Nats}, the function instance is wrapped in a Docker container for isolation.
As baseline one (henceforth referred to as \enquote{Stream}), we implement the same processing step using Apache Beam~\cite{apache2016beam} with the Google Cloud Dataflow~\cite{google2015dataflow} runner.
The Beam pipeline includes a PTransform for reading frames from Google Cloud Pub/Sub.
Preliminary experiments indicated slow job submission. 
To account for this, we greatly increase the Pub/Sub acknowledgement deadline to be longer than the experiment duration as not to miss any frames.
Next, baseline two (\enquote{FaaS}) is a Google Cloud Run Function (v2) that the load generator invokes for each frame of the stream. 
We limit the maximum instance count to one.
Lastly, baseline three (\enquote{Batch}) is also implemented as a Cloud Run Function (v2); however, the load generator first creates the entire stream and invokes the function only once with all frames as input.
As all frames need to fit into the function's memory, we intentionally configure the load generator to use the low resolution.
All systems are implemented in the Go programming language.

\paragraph{Experiment setup}
All systems run on the Google Cloud platform in region europe-west10.
The load generator and \name{} run on separate e2-standard-2 virtual machines with \SI{2}{vCPUs} and \SI{8}{GB} of memory each.
Unless specified otherwise, we use default settings for reproducibility.
For both Cloud Run Functions, we leave enough time between runs for all previous instances to be destroyed.

\paragraph{Aside} 
Note that this use case violates the fourth characteristic defined in \cref{sec:intro} in that the conversion of a stream of images to grayscale is not stateful.
This is an intentional simplification to allow the FaaS baseline to be measured without the influence of a specific storage solution for keeping state.
Furthermore, we include the cold start duration in all measurements.
We acknowledge that stream processing engines are not optimized for fast cold starts, which is precisely what want to highlight: When the arrival time is unpredictable, continuously running a stream processing engine becomes prohibitively expensive.
Consequently, a system able to deal with the characteristics defined in \cref{sec:intro} has to account for this.

\subsection{Performance metrics}

We define two performance metrics to compare \name{} against baselines one--three.
First, the \emph{cold start penalty} measures the fraction of the experiment duration that is used for the cold start, as shown in \cref{eq:csp}.
We are interested in the following three timestamps.
$t_0$ is the point in time at which the first frame is created, $t_1$ is the point in time at which the processing system receives the first frame, and $t_2$ is the point in time when the last frame was processed.
At the extremes, a $\theta$ value of 1 would signify the cold start taking up the entire processing duration.
Vice versa, as $\theta$ approaches zero, the share of time spent on the cold start does too.
Overall, the cold start penalty is a measure of how far the processing of the stream is being pushed back by a cold start.
\begin{equation}
    \theta = \frac{t_1-t_0}{t_2-t_0}
    \label{eq:csp}
\end{equation}

Next, the \emph{processing overhead} measures the difference between the entire duration ($t_2-t_0$) and the theoretically fastest possible processing duration.
This lower bound is given by the time it takes for the benchmark client to generate the stream, as --- without prior knowledge of the data --- there is no possibility of processing the stream faster than it is generated.
In our experiments, this corresponds to \SI{10}{s}, \SI{30}{s}, and \SI{60}{s}, respectively.

\subsection{Results}

\begin{figure}[t]
    \includegraphics{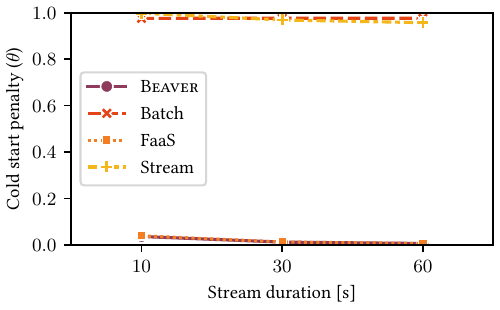}
    \caption{
        Cold start penalty for different setups and stream durations.
        Due to fast cold starts, \name{} and the FaaS approach exhibit $\theta$ values near zero.
        The batch approach improves quickly as streams get longer, at the cost of increased time until first processing.
        The stream processing approach is unfit for short streams due to slow cold starts.
    }
    \label{fig:penalty}
\end{figure}

\cref{fig:penalty} shows measurements of the cold start penalty across systems and stream durations. 
From this, two groups emerge: \name{} and FaaS exhibit $\theta$ values near zero (0.0363 and 0.0391 for \SI{10}{s} and 0.0061 and 0.0062 for \SI{60}{s}, respectively), while Stream and Batch are close to one (0.9981 and 0.9759 for \SI{10}{s} and 0.9575 and 0.9764 for \SI{60}{s}, respectively). 
With an increase of the stream duration, the Stream baseline improved the most.
The similarity between the Stream and Batch baselines indicates that, while their intention and approach to processing the data is different, their behavior was surprisingly alike.
In the Batch case, the processing step intentionally waits until the stream is completely generated and only then proceeds to process the data.
While the abstractions of the Stream baseline assume events arriving continuously, the slow cold start causes all frames to accumulate in the messaging system.
When the cold start is done, all data is already completely generated, and Stream acts as a de facto batch processing system.

\begin{figure}
    \includegraphics{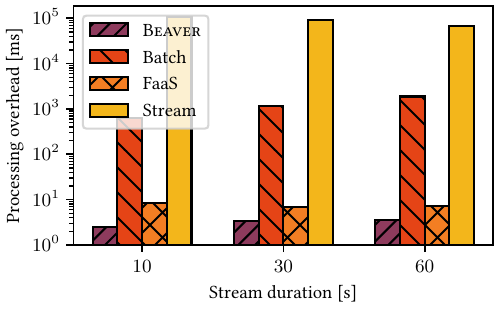}
    \caption{
        Processing overhead over theoretical minimum.
        The batch processing overhead increases for longer streams, while overheads for \name{}, FaaS, and Stream decrease slightly.
        On average, \name{} reduces the processing overhead by approximately 99~\% compared to the Stream approach and is on par with the FaaS approach.
    }
    \label{fig:overhead}
\end{figure}

\cref{fig:overhead} shows the processing overhead over the theoretical lower bound.
Here, a clear order emerges.
The Stream implementation is slowest at approximately \SI{107.84}{s}, followed by Batch at \SI{636.25}{ms}.
The FaaS and \name{} implementations perform similarly at less than \SI{10}{ms}.
The processing overheads of the FaaS and \name{} implementations stay consistent as the duration of the stream increases.
While both show cold starts in the low 100s of milliseconds, they are able to catch up to the input stream as the mean time between frames exceeds average processing time.
At the end of the input stream, when overhead becomes costly, both have warm instances ready to process the last frame with minimal overhead.
Network latency is low as both the load generator and systems under test are located in the same data center.
While \name{} outperforms FaaS slightly, this difference likely stems from the difference between production system and research prototype, as each frame needs to pass through layers of load balancers before reaching the function instance.
As expected, the Batch overhead increases with the stream duration.
Because processing only starts once the stream is complete, any increase in the amount of input data will also increase the time required for processing.
Stream improves noticeably for longer inputs.
While cold start times are similar (\SI{117.584}{s}, \SI{118.088}{s}, and \SI{121.896}{s}), we subtract the length of the stream to obtain the processing overhead.
As the input stream grows, the constant cost of starting cold is partly compensated by fast processing.
For short streams, this clearly demonstrates that large stream processing engines are not intended for our use case and a different approach is necessary.
However, the overhead improves as stream duration increases, which hints at the existence of a break-event point between systems later on.

    \section{Discussion and Outlook}\label{sec:discussion}

Our results show that serverless abstractions are well-suited for processing lightweight streams.
While serverless platforms and stream processing engines are each well optimized for their respective workloads, neither adapts gracefully to short and lightweight streams.
In turn, the behavior observed in our evaluation is a direct consequence of the assumptions embedded in these programming models.

\subsection{A sweet spot?}

We believe that stream functions address a point in the design space that has been largely unexplored.
When streams are short-running, lightweight, unpredictable, and require stateful processing, stream functions offer streaming semantics with serverless traits in a gap where FaaS falls short and for which stream processing is unoptimized.
However, stream functions are not a one size fits all solution.
Instead, they represent a complementary approach to a region in the design space for which neither current FaaS nor stream processing systems are well suited.
Our evaluation shows that, for workloads with the characteristics defined above, stream functions successfully provide streaming abstractions with the benefits of serverless computing.

We intentionally include Apache Beam with the Google Cloud Dataflow runner as a baseline because it represents the state of the art in production stream processing systems. 
Our goal is not to demonstrate a misconfiguration or inefficiency, but to evaluate how a mature streaming engine behaves when confronted with short-lived, unpredictable streams. 
The observed cold-start overhead is a direct consequence of design choices that favor throughput, fault tolerance, and long-running pipelines. 
These results therefore highlight a fundamental mismatch between existing stream processing abstractions and the workload characteristics defined in \cref{sec:intro}, rather than a limitation of Dataflow itself.

\subsection{Ordering, state, and failure semantics}

Stream functions provide execution semantics that are deliberately aligned with the lifecycle of short-lived streams.
Each incoming stream is processed by a single function invocation, which naturally preserves the arrival order of events and isolates state on a per-stream basis.
In contrast to traditional FaaS platforms, where events are scheduled independently and may be processed out of order by different instances, stream functions bind ordering and state to the duration of a stream rather than to individual events.

State within a stream function exists only for the lifetime of the stream.
If failures occur during processing, any in-memory state is lost unless explicitly externalized by the application.
While this implies weaker fault-tolerance guarantees than those offered by stream processing engines, we argue that this design choice is consistent with the targeted workload characteristics.
For short-running and lightweight streams, the cost of checkpointing and replay can easily outweigh the benefits; however, future serverless platforms could easily add this functionality if necessary.
This stands in deliberate contrast to stream processing engines, which are optimized for long-running pipelines and provide strong guarantees through mechanisms such as checkpointing, replay, and exactly-once processing.
Stream functions instead trade these guarantees for fast startup, per-stream isolation, and elastic resource usage.

\subsection{Scaling vertically}

A limitation of our design arises when the throughput of a single stream exceeds the amount a single stream function instance can process.
While stream functions scale naturally with the number of concurrent stream producers, they do not inherently support parallelizing the processing of an individual stream.

Traditional serverless functions achieve scalability through replication and load balancing of stateless function instances.
When no coordination between instances is required, events can be distributed freely without concern for ordering or shared state.
Stream functions retain this advantage across streams, but stateful processing along a single stream introduces a stricter scaling boundary.
Increasing throughput in this case requires sharding the stream across multiple function instances, which immediately raises challenges related to ordering guarantees and state consistency.
The difficulty of scaling stream functions vertically is further compounded by the diversity of stateful stream processing tasks.
For example, computing moving averages may allow for round-robin partitioning of events followed by aggregation, whereas video processing workloads may require spatial partitioning of individual frames.
These strategies differ fundamentally and cannot be derived automatically from the stream abstraction alone.
As a result, no universally applicable sharding strategy exists for stream functions.

Future serverless platforms could address this limitation in multiple ways.
One option is to restrict vertical scaling to narrowly defined cases, similar to how some existing stream processing engines parallelize operators by partitioning on explicit keys.
Alternatively, platforms could expose sharding and merging logic to developers, allowing application-specific partitioning at the cost of increased complexity.
This presents an interesting challenge that we aim to address systematically in future work.

    \section{Related Work}\label{sec:lit}

Prior work on event-driven data processing largely falls into two categories: stream processing systems and serverless computing platforms. 
Both address important aspects of scalable data processing, yet they are built around different assumptions regarding execution lifetime, state management, and elasticity. 
While a variety of works extend the serverless model in the direction of stream processing and vice versa, we argue that stream functions address a distinct area between the two fields: workloads that are naturally stream-oriented but short-running, lightweight, and unpredictable, as characterized in \cref{sec:intro}.

Stream processing systems are designed for long-running, continuous pipelines and excel at sustained, high-throughput workloads by amortizing startup costs and resource allocation over time. 
However, these systems assume that streams are continuously active and are therefore inflexible with respect to rapidly fluctuating or unpredictable workloads. 
Several efforts incorporate serverless principles into stream processing to improve elasticity, for example by using serverless functions together with streaming engines to avoid over-provisioning~\cite{song2023sponge}. 
Other work decouples operators from their streaming engines, allowing operators to be modified dynamically without restarting the entire pipeline~\cite{luthra2020operator}. 
While these approaches alleviate specific limitations, they largely preserve pipeline-based execution as the primary abstraction. 
Given their different focus, short-lived streams are not treated as first-class entities with well-defined lifetimes, state scope, and scaling behavior.

Serverless platforms, in contrast, adopt an event-at-a-time execution model that enables fine-grained elasticity and rapid scale-to-zero behavior. 
This flexibility comes at the cost of ordering guarantees and seamless inter-event logic, as platforms provide neither control over which instance processes a given event nor direct addressability of function instances~\cite{hellerstein2019twostepsserverless}. 
A substantial body of work addresses these limitations by equipping functions with key--value stores or shared logs to retain state across invocations~\cite{jia2021boki,sreekanti2020cloudburst,shillaker2020faasm,pfandzelter2023enoki,barcelonapons2019faastrack} or by enhancing coordination and communication between instances through scheduling, function composition, or explicit messaging~\cite{pu2019shuffling,sadjad2017excamera,dukic2020photons,schirmer2024fusionizepp,kowallik2026konflux,akkus2018sand,copik2023fmi}.
In addition, richer trigger mechanisms can reduce the need for explicit state management by allowing functions to react to multiple related events at once, offering a complementary approach to stateful execution~\cite{lopez2020triggerflow,carl2025met}. 
Despite these advances, existing systems retain independently scheduled function invocations as their core abstraction. 
Consequently, they scale individual events rather than streams and provide limited support for expressing inter-event logic over continuous data flows of unknown duration.

    \section{Conclusion}\label{sec:conclusion}

In this paper, introduced stream functions as a programming abstraction for a class of workloads that lies between traditional stream processing and serverless computing. 
We target applications that are stream-oriented yet short-running, lightweight, and unpredictable --- characteristics that violate the assumptions of both models in different ways. 
Stream functions combine streaming semantics with the elastic lifecycle management of serverless platforms, treating a short-lived stream itself as the unit of execution, state, and scaling. 
By aligning execution and state lifetime with the duration of a stream, we argue, stream functions provide a natural foundation for processing ephemeral streams.

Our evaluation with \name{}, our proof-of-concept implementation, demonstrates that this abstraction can substantially reduce overhead for short-running streams. 
Compared to a mature stream processing system, \name{} reduces processing overhead by \textasciitilde{}\SI{99}{\percent} while retaining comparable performance to event-at-a-time serverless execution. 
These results highlight that the inefficiencies observed in existing systems are not primarily implementation artifacts, but stem from a fundamental mismatches between workload characteristics and the underlying programming model.
We believe that stream functions highlight the importance of reconsidering execution granularity in serverless systems. 
As cloud applications increasingly combine event-driven and streaming behavior, abstractions that bind execution and state to meaningful units of work are essential. 
Stream functions represent a first step toward such abstractions and open up new opportunities for building efficient, elastic, and expressive data processing systems.

    \bibliographystyle{ACM-Reference-Format}
    \bibliography{references}

\end{document}